%% file: main.tex
\documentclass[sigconf]{acmart}
\renewcommand\footnotetextcopyrightpermission[1]{}

\usepackage{graphicx} 

\title{\textit{SpecEval}: Evaluating Code Comprehension in Large Language Models via Program Specifications}

\newcommand{\ourname}[0]{\textit{SpecEval}\xspace}

\newcommand{\lz}[1]{\textcolor{black}{#1}}

\usepackage{enumerate}
\usepackage{enumitem}
\usepackage{xspace}
\usepackage{subcaption}
\usepackage{booktabs}
\usepackage{multirow}
\usepackage{multicol}
\usepackage{tcolorbox}
\begin{document}


\author{Lezhi Ma}
\email{lezhima@hotmail.com}
\affiliation{
  \institution{Nanjing University}
  \city{Nanjing}
  \state{Jiangsu}
  \country{China}
}

\author{Shangqing Liu}
\email{shangqingliu@nju.edu.cn}
\affiliation{
  \institution{Nanjing University}
  \city{Nanjing}
  \state{Jiangsu}
  \country{China}
}

\author{Lei Bu}
\email{bulei@nju.edu.cn}
\affiliation{
  \institution{Nanjing University}
  \city{Nanjing}
  \state{Jiangsu}
  \country{China}
}

\author{Shangru Li}
\email{shangruli1013@gmail.com}
\affiliation{
  \institution{Nanjing University}
  \city{Nanjing}
  \state{Jiangsu}
  \country{China}
}

\author{Yida Wang}
\email{wangyida2002@hotmail.com}
\affiliation{
  \institution{Nanjing University}
  \city{Nanjing}
  \state{Jiangsu}
  \country{China}
}

\author{Yang Liu}
\email{yangliu@ntu.edu.sg}
\affiliation{
  \institution{Nanyang Technological University}
  \city{Singapore}
  \country{Singapore}
}

\begin{abstract}
Large Language models (i.e., LLMs) have achieved impressive performance in automated software engineering. Extensive efforts have been made to evaluate the abilities of code LLMs in various aspects, with an increasing number of benchmarks and evaluation frameworks proposed. Apart from the most sought-after capability of code generation, the capability of code comprehension is being granted growing attention. Nevertheless, existing works assessing the code comprehension capability of LLMs exhibit varied limitations. Evaluation frameworks like CRUXEval and REval usually focus on code reasoning tasks over a certain input case, leading to a limited range of execution traces covered, resulting in a loss in code semantics examined and the inability to assess the comprehensive understanding of LLMs concerning the target program.

To tackle these challenges, we propose \textit{SpecEval}, a novel black-box evaluation framework to evaluate code comprehension in LLMs via program specifications. Inspired by the idea that specifications can act as a comprehensive articulation of program behaviors concerning all possible execution traces, we employ formalized program specifications to represent program semantics and perform comprehensive evaluations. In particular, four specification-related tasks are designed meticulously to assess the capability of LLMs from basic to advanced levels. Counterfactual analysis is further conducted to study the performance variance of LLMs under semantics-preserving perturbations. Systematic experiments are conducted on six state-of-the-art LLMs. Extensive experimental results present a below-satisfactory performance of LLMs on specification-related tasks, revealing the limitations of existing LLMs in terms of articulating program semantics with formal specifications. Counterfactual analysis also reveals the sensitivity of LLMs towards semantic-preserving perturbations.

\end{abstract}

\maketitle
\pagestyle{plain} 

\input{sections/introduction}
\input{sections/motivation}
\input{sections/approach}
\input{sections/setup}

\input{sections/evaluation}

\input{sections/discussion}

\input{sections/related}
\input{sections/conclusion}


\bibliography{ref}
\bibliographystyle{plainnat}

\end{document}

%% file: sections/introduction.tex
\vspace{-2mm}
\section{Introduction} \label{sec:introduction}
Automated software engineering, encompassing areas such as code generation~\cite{liu2024your,chen2021evaluating}, program repair~\cite{xia2023universal,Nilizadeh2021AprFormalMethods, kong2024contrastrepair}, and code summarization~\cite{liu2020retrieval,mcburney2015automatic}, has persistently remained a prominent research focus in both academic and industrial practice. Early research in this field adopted traditional techniques, such as template-based~\cite{flanagan2001houdini,ernst2007daikon} and retrieval-based~\cite{hayati2018retrieval,liu2020retrieval} methods, to address these issues. However, due to technology and data availability constraints, these methods often struggled to deliver satisfactory performance. With the rise of deep learning, techniques including RNN~\cite{yu2019review}, Transformer~\cite{vaswani2017attention}, and GNNs~\cite{wu2020comprehensive} have been applied to this field. Extensive research on program semantics learning with advanced neural network models drives significant advancements in various SE applications. Program semantics are the essence of a program, programs with identical semantics may have different expressions. Consequently, numerous studies~\cite{susanto2017semantic,zhou2019devign,peng2021could} have explored different techniques, including tree-based and graph-based approaches, to improve the neural networks in \textit{learning program semantics}.

Large language models (LLMs)~\cite{openai2023gpt,achiam2023gpt,wei2023magicoder,nijkamp2023codegen2,lozhkov2024starcoder,guo2024deepseek,touvron2023llama,roziere2023code,guo2025deepseek} have recently impressed the community with their outstanding performance, vastly surpassing previous techniques in various software engineering tasks. Meanwhile, a wide range of software development tools~\cite{github2024copilot,jetbrains2024ai} powered by these models have significantly increased the productivity of software developers. Compared to earlier techniques such as graph-based ones that explicitly incorporated code semantics into model training to help the model comprehend program semantics, LLMs usually treat the code snippet as pure text and rely on the straightforward pre-training technique, i.e., next-token prediction, to train on a massive amount of code corpus. This raises a question: does next-token pertaining technique truly enable models to \textit{comprehend} the complex code semantics?


Numerous benchmarks, such as HumanEval~\cite{chen2021evaluating}, MBPP~\cite{austin2021program}, and SWE-bench~\cite{jimenez2023swe}, are proposed to evaluate the coding capacity of LLMs. Although these benchmarks can, to some extent, reflect the code generation capabilities of different models, using the criterion of whether the generated code passes test cases is still insufficient to determine whether the models genuinely understand the code~\cite{yang2024unveiling}. 
For more in-depth analysis, CRUXEval~\cite{gu2024cruxeval} and REval~\cite{chen2024evaluating} define certain program execution tasks, such as code coverage prediction and output prediction, to assess the code reasoning capabilities of LLMs by comparing the model's generated results with actual program runtime outcomes. Nevertheless, the dynamic execution results of a program depend on its input, making it challenging to ensure that a comprehensive set of inputs can cover all possible execution traces~\footnote{The execution trace refers to the transitions of program states~\cite{plotkin1981structural} where the states are defined as the set of all program variables including their types and values.}. This leads to \textit{limited} program inputs invoking \textit{limited} execution traces for a program and cannot comprehensively reflect the full program semantics. Some probing techniques~\cite{hernandez2022ast, wan2022they, ma2024unveiling} are adopted to analyze the code understanding, such as code syntax and semantic knowledge. They typically require access to a model's intermediate vector representations for corresponding inputs to investigate whether the model has learned specific attributes. As these techniques usually require direct access to the model, they are not feasible for black-box models such as GPT-3.5 and GPT-4.


These challenges inspire us to explore other forms of representation for program semantics, which are required to articulate program behaviors in multiple granularities. Formalized program specifications encompass precise claims that describe the intended or actual program behaviors in \textit{formal} languages. By specifying the constraints on program variables or the transition of program states, the strong program specifications can effectively articulate program semantics, accurately capturing program behaviors either generally or in detail. Moreover, the correctness of formal specifications can be automatically and effectively validated by corresponding specification verifiers~\cite{cok2011openjml,ahrendt2014key,kirchner2015frama}, overcoming the limitations of natural language specifications such as code summaries.

Inspired by these ideas, we propose \ourname, a novel black-box evaluation framework featuring the representation of program semantics in the form of formal program specifications. We define multiple specification-related tasks for subject LLMs to complete. These tasks require LLMs to demonstrate an understanding of code by summarizing code behavior into formalized statements. Leveraging automated procedures for determining the correctness of specifications, the assessment can be conducted rigorously and efficiently. Specifically, four specification-related tasks are designed, including \textit{Judgement}, \textit{Selection}, \textit{Infilling}, and \textit{Generation}. In the \textit{Judgement} task, LLMs are required to judge the correctness of the given specification. The \textit{Selection} task involves selecting the most appropriate candidate specification for the target program. The \textit{Infilling} task requires LLMs to complete partially provided specifications by filling in the correct sub-expressions. Finally, the \textit{Generation} task challenges LLMs to generate correct specifications entirely from scratch. For each task, we further introduce \textit{counterfactual analysis} to cast code comprehension as the problem of determining how controlled input code changes result in model output changes by the semantic-preserving perturbations, with mutation operators including \textit{Def-use Break}, \textit{If-else Flip}, \textit{Independent Swap}, \textit{Name Random}, and \textit{Name Shuffle}. 
For evaluation, we collect \textbf{204} Java programs with verifiable ground-truth specifications in JML~\cite{leavens1998jml} style from existing datasets. Test cases for these programs are also generated to assist the specification validation process. We conduct experiments on six large language models, including state-of-the-art code LLMs and general LLMs, for a systematic comparison between the popular models. Experimental results show that LLMs still have a long way to go before being able to fully articulate program semantics with formal specifications, with less than satisfactory performance in most specification-related tasks. Furthermore, Counterfactual analysis discloses the sensitivity of LLMs to different extents towards semantic-preserving perturbations. 

In summary, the main contributions of our work include:
\vspace{-1mm}
\begin{itemize} [leftmargin=*]
    \item A novel black-box evaluation framework assessing the program semantics learning abilities of LLMs, featuring the representation of program semantics in the form of formalized specifications. Four different tasks and Counterfactual analysis are designed to study the performance of LLMs from multiple aspects.
    \item An adapted benchmark based on existing datasets~\cite{sosylab2024svcomp,github2024framac,Nilizadeh2021AprFormalMethods,ma2024specgen}, consisting of 204 Java programs along with five semantics-equivalent variants, as well as ground-truth specifications.
    \item A thorough evaluation of six state-of-the-art code LLMs and general LLMs, revealing the limitations of LLMs concerning the expression of program semantics with formal specifications, underscoring the directions for future enhancement of LLMs.
\end{itemize}

%% file: sections/motivation.tex
\section{Background and Motivation} \label{sec:background}

\subsection{Program Specification} \label{sec:background_specification}
Program specification refers to precise statements that define a program's intended or actual behaviors.  According to the language adopted, program specifications can be categorized into natural language specifications and formalized specifications. Natural language specifications primarily encompass software documentation and code comments. Another significant portion of program specifications is articulated using formal languages, such as mathematical expressions~\cite{leavens1998jml,kirchner2015frama} to specify the constraints on program behaviors, and automata~\cite{lo2006quark,shoham2007static} to describe the transitions of program states. Formalized specifications have been widely adopted in automated software engineering tasks, such as requirement engineering~\cite{aman2013reverse,sharifi2020symboleo}, software testing~\cite{Mesbah2012InvariantBasedTesting,nguyen2023tc4mt}, and model checking~\cite{cabodi2008strengthening,beyer2015boosting}.

\begin{figure}[!t]
    \centering
    \includegraphics[width=0.45\textwidth]{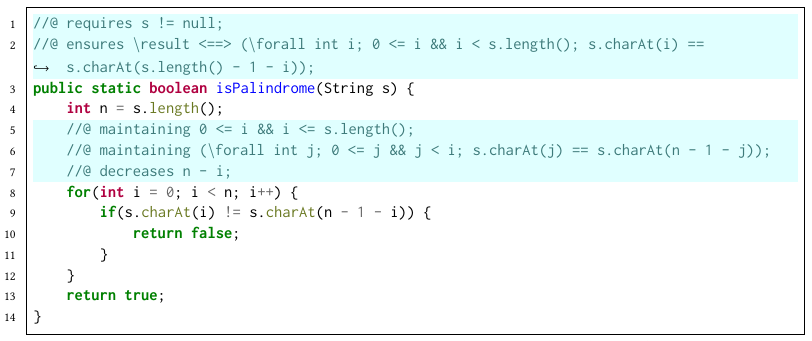}
    \vspace{-3mm}
    \caption{An example program and corresponding ground-truth JML specification (highlighted in blue).}
    \label{fig:spec_example}
    \vspace{-6mm}
\end{figure}

In \ourname, we mainly focus on formal specifications written in Java Modeling Language (JML)~\cite{leavens1998jml} describing the actual behaviors of Java programs. One example program \texttt{isPalindrome} and corresponding ground-truth specifications are illustrated in Fig.~\ref{fig:spec_example}. The program aims to check whether a given input string \texttt{s} is a palindrome string, i.e., a string that reads the same forward and backward. The specifications are instrumented within the program in the form of comments starting with \texttt{//@}. Generally, the specifications involved can be divided into three categories: Preconditions (Line 1), Postconditions (Line 2), and Loop Invariants (Line 5, 6, and 7).
Together, the specifications mentioned above constitute a detailed articulation of the semantics of the target program. 


It is worth noting that formalized program specifications are more rigorous, eliminating ambiguity compared with specifications in natural language. Moreover, formalized program specifications can thoroughly articulate program semantics. Program semantics essentially lies in the evolution of program states, which are usually defined by the set of variable values. By specifying the constraints on program variables or the transition of program states, strong program specifications are rich in semantic information and can accurately capture program behaviors. In addition, the correctness of formal specifications can be effectively validated by specification verifiers~\cite{cok2011openjml,ahrendt2014key,kirchner2015frama}, which can automatically check the consistency between the given specifications and the target program. Consequently, compared with natural language specifications, the correctness of program specifications can be automatically verified.



\vspace{-2mm}
\subsection{Motivation}
Program semantics are the core of a program, and only when code models can effectively learn these semantics can they provide reliable support and assurance in software development. In light of the critical role of code semantics, a substantial body of research has attempted to evaluate whether LLMs genuinely comprehend code through various analytical techniques. 
A typical white-box technique involves probing analysis~\cite{hernandez2022ast, wan2022they, ma2024unveiling} to explore whether code models have learned specific code attributes from the trained code datasets. It leverages the transformation of code semantic information, such as data flow, into high-dimensional vector representations in the latent space, enabling an examination of whether the model truly understands the semantics of the code. Another typical black-box framework REval~\cite{chen2024evaluating} evaluates the code reasoning ability of LLMs by inferring the program runtime behaviors. Inspired by CRUXEval~\cite{gu2024cruxeval}, REval designs four evaluation tasks: Code Coverage Prediction requires LLMs to predict whether a statement will be executed given a certain input, Program State Prediction requires LLMs to reason about the value of a specific variable, Execution Path Prediction requires LLMs to predict the next statement to be executed, and Output Prediction requires LLMs to generate the program's output directly from the given input.

However, existing works face various limitations. Probing analysis usually requires access to the model's intermediate vector representations for analysis, which is not feasible for black-box commercial models such as GPT-3.5 and GPT-4. As for the black-box approach REval, it should be noted that in each of the four tasks involved, LLMs are provided with one specific input respectively. For a deterministic program, one specific input can only result in one particular trace of program state transition, i.e. the execution trace. This observation leads to the fact that for each independent task assigned, LLMs are only reasoning about program behaviors within \textit{one specific} execution trace, whereas there exists an infinite number of execution traces (corresponding to the infinite number of inputs) within the program state space other than the one being examined. Nevertheless, program semantics is not only about the program behaviors during one specific execution but also about the general pattern of program state transition adaptable to all potential program states. The set of program behaviors for all potential program states and execution traces constitutes full program semantics. This indicates that the practice employed by REval could cause a loss in the program semantics examined, since the evaluation tasks and inputs are limited, leaving the other program behaviors in an \textit{infinite} number of execution traces uncovered.

To tackle the issues mentioned above, a new carrier to articulate program semantics is necessary. As described in Section~\ref{sec:background_specification}, specifications can thoroughly articulate the program semantics. Compared to the perspectives of existing works, specifications can describe the abstract patterns between program variables that should be adaptable to all possible execution traces. For instance, the post-condition on Line 2 illustrated in Fig.~\ref{fig:spec_example} articulates a precise constraint on the input string \texttt{s} and the return value \texttt{\textbackslash result}. The constraint is correct, yet \textit{general}, i.e., all pairs of possible input strings and corresponding return values must satisfy this constraint. Because the specification works for all possible inputs, it acts as an abstract pattern for all possible execution traces since each valid execution trace must be triggered by a specific input. Proceeding from this idea, we propose \ourname, a black-box evaluation framework for LLMs featuring the representation of program semantics in the form of formal program specifications. The defined specification-related tasks can evaluate the comprehension of LLMs concerning the comprehensive program semantics. By requesting LLMs to complete these tasks, their capabilities to comprehend code can be evaluated by assessing the quality of the generated specifications. 

%% file: sections/approach.tex
\vspace{-1mm}
\section{Methodology} \label{sec:approach}


In this Section, we aim to present a detailed introduction to our framework. We first describe the evaluation tasks in our framework. For each task, we introduce the \textit{counterfactual analysis}, framing code comprehension as the challenge of assessing how controlled input code modifications, through semantic-preserving perturbations, lead to changes in the model's output.
This framework enables a comprehensive assessment of the code comprehension capabilities of LLMs, ranging from superficial to deeper levels of understanding.



\vspace{-2mm}
\subsection{Evaluation Tasks and Metrics} \label{sec:tasks}

We designed four tasks related to formal program specifications. These tasks not only request LLMs to summarize abstract semantic information into formalized statements but also require them to comprehend the given specifications and take appropriate actions based on their understanding. \lz{Some tasks directly correspond to specific software engineering techniques, while others implicitly assess the LLMs' ability to comprehend formal languages.}

\vspace{-1mm}
\subsubsection{Specification Correctness Judgement} \label{sec:task_judge}
Specifications possess \textit{correctness}, defined by their consistency with program behaviors. A correct specification is about constraints that \textit{always hold} for all program states in its scope. We denote the \textbf{actual correctness} of the candidate specification $s$ as a function $\mathbf{Cr}: S \rightarrow \{\mathbf{true},\mathbf{false}\}$, which takes a value of $\mathbf{Cr}(s)=\mathbf{true}$ if $s$ is correct and $\mathbf{Cr}(s)=\mathbf{false}$ otherwise. In this task, subject LLM is exploited to decide the correctness of a given candidate specification on a target program. \lz{This task closely aligns with software model checking~\cite{jhala2009software}, a static technique to prove whether a program satisfies specific properties.} The correct candidates are extracted from the ground-truth specifications of the target program. The incorrect candidates are obtained by modifying essential components within the ground-truth specification, including replacing variable names with other variables in the same scope, altering variable operators, and swapping statement predicates (i.e., between \texttt{\textbackslash forall} and \texttt{\textbackslash exists}). Each component of interest is decided randomly whether it will be modified, with equal probabilities (at 50\%) of being modified or staying unchanged. At least one modification is performed to ensure the specification does not stay the same. \lz{Runtime checking is performed on the mutated candidates to ensure their incorrectness. A candidate should be mutated again if it remains correct after mutation.}

\begin{itemize} [leftmargin=*]
    \item \textbf{Task Formalism.} Given a target program $P$, and a candidate specification $s$, a judgment task $t \in T$ is a 2-tuple $(P,s)$, where the subject model $\mathcal{M}: T \rightarrow \{\mathbf{true},\mathbf{false}\}$ is required to decide whether the given specification is correct for the target program.
    \item \textbf{Evaluation Metrics.} In this task, \textit{Accuracy} measures the percentage of correctly judged specifications. The \textit{Accuracy} for this task can be calculated by the following formula:
    $$
    Accuracy = \frac{1}{|T|}\sum\limits_{(P,\ s) \in T}\mathbb{I}(\mathcal{M}(P,s)=\mathbf{Cr}(s))
    $$
    where $\mathbb{I}:\{\mathbf{true},\mathbf{false}\}\rightarrow\{0,1\}$ is the indicator function that maps $\mathbf{true}$ to 1 and $\mathbf{false}$ to 0.
\end{itemize}

\vspace{-1mm}
\subsubsection{Specification Candidates Selection} \label{sec:task_select}
Specifications possess varying degrees of \textit{strength}. While weaker specifications such as \texttt{ensures true} can still pass the verification and should be considered correct, they often describe only trivial information about the program. In contrast, stronger specifications provide a more comprehensive expression of the program's semantics. Therefore, in this task, we consider both the correctness and strength when constructing candidates for the given target program. In particular, four candidate specifications are provided to the model, which is required to choose the \textit{most appropriate} (i.e., correct and strongest) one for the given target program. The most appropriate candidate (denoted as $\hat{s}$) in four candidates is extracted from the ground-truth specifications for the target program. Other candidates can be either modified specifications or trivial specifications. The former is obtained by modifying the ground-truth specification with the same logic mentioned in Section~\ref{sec:task_judge}. The latter is randomly chosen from a set of pre-defined simple specifications that are too weak to be incorrect, such as \texttt{ensures \textbackslash result <= Integer.MAX\_VALUE}. At most one trivial specification is included in each task. \lz{This task is analogous to \textit{Code Preference Prediction}~\cite{liu2024learning}, where LLMs are required to choose the most appropriate code snippet according to given requirements. The difference is that LLMs should choose between the specifications here instead of the code.}
\begin{itemize} [leftmargin=*]
    \item \textbf{Task Formalism.} Given a target program $P$ and a set of four candidate specifications $S=\{s_1,s_2,s_3,s_4\}$, a selection task $t \in T$ is a 2-tuple $(P,S)$, where the subject model $\mathcal{M}: T \rightarrow S$ should return the \textit{most appropriate} candidate within $S$ for $P$. The ground truth for this task is the correct candidate, denoted as $\hat{s}$.
    \item \textbf{Evaluation Metrics.} In this task, \textit{Accuracy} measures the percentage of correctly selected specifications. The \textit{Accuracy} for this task can be calculated by the following formula.
    $$
    Accuracy = \frac{1}{|T|}\sum\limits_{(P,\ S) \in T}\mathbb{I}(\mathcal{M}(P,S)=\hat{s})
    $$    
\end{itemize}

\subsubsection{Specification Infilling}
In this task, an uncompleted specification (partially masked with a placeholder) is provided to the subject LLM, which is required to fill in the placeholder with appropriate sub-expressions according to the given target program so that the final specification can pass the verifier. The uncompleted specification is acquired by randomly masking specific components of the ground-truth specifications with placeholders. Only one placeholder is inserted for each task. The components of interest include array index, variable and method names, and boundary constraints of \texttt{\textbackslash forall} and \texttt{\textbackslash exists} statements, which can be identified by parsing specifications into Abstract Syntax Trees. \lz{The task formulation is highly similar to \textit{code infilling}~\cite{fried2022incoder}, a derived coding task widely adopted in various LLM benchmarks~\cite{bavarian2022efficient,gong2024evaluation}, except that it is the specification that is to be infilled here rather than the code.}
\begin{itemize} [leftmargin=*]
    \item \textbf{Task Formalism.} Given a target program $P$, and an uncompleted specification $s$, an infilling task $t \in T$ is a 2-tuple $(P,s)$, where the subject model $\mathcal{M}: T \rightarrow S$ is required to fill in the placeholder of $s$ according to the target program and return the filled-in version.
    \item \textbf{Evaluation Metrics.} In this task, we use \textbf{\#Pass} and \textit{Accuracy} for measurement. \textbf{\#Pass} denotes the number of programs $P$ for which the specification infilled by model $\mathcal{M}$ are correct. \textit{Accuracy} measures the percentage of correctly infilled specifications. The \textit{Accuracy} for this task can be calculated as:
    $$
    Accuracy = \frac{1}{|T|}\sum\limits_{(P,\ s) \in T}\mathbb{I}(\mathbf{Cr}(\mathcal{M}(P,s)))
    $$    
\end{itemize}

\subsubsection{Specification Generation} \label{sec:task_generation}
In this task, the subject LLM is provided with a program $P$ with no instrumented specifications. The LLM is exploited to generate a set of specifications $S$ from scratch, describing the general behaviors and functionalities of the target program as closely as possible. The generated specifications should include pre/post-conditions for all methods and loop invariants for all loops involved in the target program. \lz{Such task formulation is frequently adopted in recent works on program specification generation techniques~\cite{ma2024specgen,wen2024enchanting}.}
\begin{itemize} [leftmargin=*]
    \item \textbf{Task Formalism.} Given a program $P$, the model $\mathcal{M}: \mathbb{P} \rightarrow 2^S$ should generate an appropriate set of specifications $S$ for $P$.
    \item \textbf{Evaluation Metrics.} In this task, \textit{Precision} and \textit{Recall} are adopted to measure the performance of LLMs. \textit{Precision} measures the percentage of correct specifications that can pass the verifier over all the generated specifications. \textit{Recall} measures the percentage of correctly generated \textbf{ground-truth} specifications (i.e., correct and strongest) over all the ground-truth specifications for the program $P$. The above metrics can be calculated by
    $$
    Prec(P) =\frac{
                    \sum_{s \in \mathcal{M}(P)}
                    \mathbb{I}(\mathbf{Cr}(s))
                }
                {|\mathcal{M}(P)|} \quad
    Rec(P)=\frac{
                \sum_{\hat{s} \in \hat{S}_P}
                \mathbb{I}(\hat{s} \in \mathcal{M}(P))
            }
            {|\hat{S}_P|}
    $$
    where $\hat{S}_P$ denotes the set of all ground-truth specifications for program $P$.
    Apart from precision and recall, we also use \textbf{\#Pass} metric to denote the number of programs $P$ for which all specifications generated by model $\mathcal{M}$ can pass the verifier.
\end{itemize}
\lz{It is noteworthy that calculating \textit{Recall} necessitates determining whether each ground truth specification is covered by the model's outputs, i.e., whether $\hat{s} \in \mathcal{M}(P)$ holds. This involves determining the equivalence between a ground truth specification $\hat{s} \in \hat{S}_P$ and each of the model-generated specifications $s \in \mathcal{M}(P)$. To address this issue, we utilize a bidirectional implication operator to combine the two specifications into a new composite specification $\hat{s} \Leftrightarrow s$. The equivalence between the two specifications is then determined by evaluating the correctness of this composite specification. Thus, if there exists a model-generated specification $s \in \mathcal{M}(P)$ that is equivalent to $\hat{s}$, we can conclude that $\hat{s} \in \mathcal{M}(P)$ and vice versa.}

\vspace{-2mm}
\subsection{Counterfactual Analysis} \label{sec:counterfactual}
Counterfactual Analysis is a technique conducted by observing the variance in the model performance after changing the model inputs in a particular way~\cite{verma2020counterfactual}. In \ourname, we respectively adopt several typical semantic-preserving perturbations to analyze the performance variance of LLMs on each kind of the evaluation tasks. 

\subsubsection{Semantic-preserving Perturbations} \label{sec:perturbations}

\begin{figure*}[!t]
    \centering
    \includegraphics[width=1\textwidth]{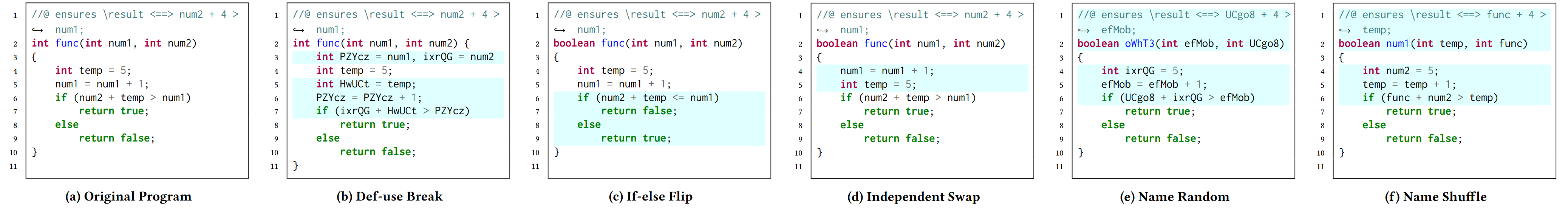}
    \vspace{-6mm}
    \caption{Illustrations for the semantic-preserving perturbations, including the original program, corresponding specifications, and all perturbed programs. The code and specifications modified by the perturbations are highlighted in blue.}
    \label{fig:perturbations}
    \vspace{-4mm}
\end{figure*}

\lz{Among the common semantically equivalent program transformation strategies~\cite{hooda2024large,le2024evaluating}, we adopted the following five general ones as the perturbations.}
\begin{itemize} [leftmargin=*]
    \item \textit{Def-use Break}. Def-use chain refers to the relationship between the definitions of a certain variable (where it is assigned with a value, e.g., \texttt{x = 0}) and its subsequent uses (where its value is accessed, e.g., \texttt{func(x)}). The \textit{Def-use Break} perturbation aims to break these chains within target programs by assigning the value of a variable \texttt{var1} to a newly added variable \texttt{var2} and altering all subsequent uses of \texttt{var1} to \texttt{var2}. In our implementations, the variable name for \texttt{var2} consists of five random letters or digits. For instance, in the program shown in Fig.~\ref{fig:perturbations}, a new variable \texttt{PZYcz} inherits the value of variable $\texttt{num1}$, and replaces the latter in the remaining part of the program. Similar modifications are performed on variable \texttt{num2} and local variable \texttt{temp}. Since the new variable completely supersedes the role of the original variable, the perturbation is guaranteed to be semantic-preserving.
    \item \textit{If-else Flip}. This perturbation swaps the branches of an if-else statement. This is done by replacing the content of the two branches with each other and negating the branch condition.
    Such perturbation does not affect the program semantics. 
    \item \textit{Independent Swap}. For ease of handling, we adopt a \textit{safe} definition for independent statements. For two adjacent statements $S_1$ and $S_2$ in the same basic block, denote the sets of variables defined and used by statement $S_i$ as $V^{def}_i$ and $V^{use}_i$ respectively. If the following three conditions are satisfied simultaneously: (1) $V^{def}_1 \cap V^{def}_2=\emptyset$ (2) $V^{use}_1 \cap V^{def}_2=\emptyset$ (3) $V^{def}_1 \cap V^{use}_2=\emptyset$, then the two statements are considered \textit{independent} from each other and can be swapped. For instance, in the original program shown in Fig.~\ref{fig:perturbations}, statements on line 4 and line 5 constitute a pair of independent statements. Swapping such pairs of statements will not disrupt the original data flow and is semantic-preserving.
    \item \textit{Name Random}. This perturbation assigns randomly generated names to all program variables and methods. Consistent with \textit{Def-use Break}, the generated names consist of five random letters or digits. Altering the variable name of a program does not affect program semantics, given that the altered variable names do not conflict with others. The corresponding variable names in the specifications also need to be changed accordingly.
    \item \textit{Name Shuffle}. The idea of this perturbation is the same with \textit{Name random} but is done by reshuffling all the variable names within the original program rather than generating new names.  The corresponding specifications are changed accordingly. Similarly, the perturbation does not affect the original program semantics.
\end{itemize}
\vspace{-1mm}
It is worth noting that for some of the perturbations, the specifications involved in the original program should be mutated simultaneously. For instance, \textit{Name Random} and \textit{Name Shuffle} alter all the variable names in the original program, where the specifications should undergo identical lexical modifications since the variable names are also involved in the specifications. \textit{If-else Flip} perturbation may change the positions of some specifications within the target if-else branch, so the specifications involved need to be migrated. Such modifications to specifications pose no impact on their semantics. Furthermore, task prompts involving specifications must also be mutated to keep them equivalent to the original tasks, such as the candidate specifications in \textit{Judgement} and \textit{Selection} tasks.

\vspace{-1mm}
\subsubsection{Evaluation Metrics}
We adopt different metrics for different tasks in counterfactual analysis. For \textit{Judgement}, \textit{Selection}, and \textit{Infilling}, we adopt \textit{Jaccard Distance} to measure the impact of perturbations on the task as
$J = 1 - |\mathbb{P}\cap\mathbb{P}^\prime| / |\mathbb{P}\cup\mathbb{P}^\prime|$
, where $\mathbb{P}$ and $\mathbb{P'}$ refer respectively to the set of \textit{original} and \textit{perturbed} programs that an LLM can successfully handle for one kind of perturbation in Section~\ref{sec:perturbations}. The programs in $\mathbb{P}\cap\mathbb{P}^\prime$ refer to the LLM successfully handles both the original and the perturbed versions, indicating that its performance is \textit{consistent} on these programs, without being interfered by the perturbation, whereas other programs in $\mathbb{P}\cup\mathbb{P}^\prime$ witness a varied performance on different versions. The \textit{Jaccard Distance} measures the impact of perturbations by calculating the percentage of \textit{inconsistently-performing} programs. A high distance indicates a strong impact that the corresponding perturbation poses.

For the \textit{Generation} task, since it is quantitatively measured by \textit{precision} and \textit{recall}, we define \textit{Average Variance} to measure the impact of perturbations for this task. For a specific type of perturbation in Section~\ref{sec:perturbations}, the \textit{Average Variance} of metric $m$ where $m \in \{Precision, Recall\}$ can be calculated as
\vspace{-1mm}
$$
v_{m} = \frac{1}{|\mathbb{P}|} \sum\limits_{(P,P^\prime) \in \mathbb{P}} |m(P^\prime)-m(P)|
$$
where $\mathbb{P}=\{(P,P^\prime)\ |\ m(P)>0 \vee m(P^\prime)>0\}$, $(P,P^\prime)$ denotes pairs of the original and the perturbed programs. To keep consistent with the \textit{Jaccard Distance}, we only consider $(P,P^\prime)$ where either $P$ or $P^\prime$ achieve a non-zero value of $m$. The average variance of precision and recall is denoted as $v_{prec}$ and $v_{rec}$, respectively.

%% file: sections/setup.tex
\vspace{-1mm}
\section{experimental setup} \label{sec:setup}

\subsection{Benchmark Construction}

We construct a benchmark that includes programs and corresponding ground-truth specifications for evaluation. Concerning the specification language, we adopt the Java Modeling Language (JML)~\cite{leavens1998jml} for Java programs for evaluation since they can be conveniently validated by OpenJML~\cite{cok2011openjml}. We use the released datasets by the latest two works on LLM-based specification generation~\cite{wen2024enchanting,ma2024specgen}. One work by Ma et al.~\cite{ma2024specgen} utilizes two datasets, \textbf{SpecGenBench} and \textbf{the Java Category of SV-COMP Benchmark}~\cite{sosylab2024svcomp}. The former contains 120 Java programs with manually-written strong specifications and the latter is widely adopted in the evaluation of software verification tools. For \textbf{SpecGenBench}, we extract 117 programs from the dataset along with corresponding specifications, whereas 3 programs are discarded due to incompatibility with OpenJML. For \textbf{the Java Category of SV-COMP Benchmark}, as these programs are without corresponding specifications, we manually craft them following the similar procedure adopted in the construction of SpecGenBench~\cite{ma2024specgen}. Extensive efforts have been conducted to manually write these specifications, and we have succeeded in 47 programs from SV-COMP that are compatible with OpenJML, and the written specifications can fully specify the program semantics. Another work~\cite{wen2024enchanting} leverages another dataset, \textbf{Frama-C-Problems}~\cite{github2024framac}, which contains 47 C programs and corresponding ACSL~\cite{baudin2008acsl} specifications. To cater to Java style, we manually rewrote the programs in Java and altered the specifications to JML-style. 40 programs that can be equivalently translated into Java programs. Others are discarded due to specific features (i.e., pointers) of the C programming language that cannot be trivially translated.

All specifications have been cross-validated by three experts in formal methods and programming, where the semantics of the specifications and corresponding programs are compared. The specifications will be re-formulated until their strength is confirmed by all the experts to ensure their quality and ability to describe the behaviors and functionalities of the target program \textit{thoroughly}. Additionally, all specifications are also ensured to be \textit{verifiable}, i.e., able to pass the verification of the verifier (namely OpenJML~\cite{cok2011openjml} in our work), so the correctness of the specifications can be guaranteed. \lz{Eventually, 204 programs are collected for evaluation. The average LoC (Lines of Code) and CC (Cyclomatic Complexity) of the programs adopted are 20.06 and 6.34, respectively. Among the programs, 36 are sequential (i.e., no branches or loops), 77 are loop-free but with branches (i.e., if-else or switch statements), 70 have single layer loops, and 20 have nested loops.}

During the experiment, we adopt runtime checking supported by specification verifiers to validate the correctness of the specifications generated by the studied LLMs (i.e., to calculate $\mathbf{Cr}(s)$ mentioned in Section~\ref{sec:tasks} for a given specification $s$). A specification is considered correct (i.e., $\mathbf{Cr}(s) = \mathbf{true}$) if it maintains its consistency with the target program during a runtime execution process. To this end, test cases are also prepared for each program to invoke their execution. Similar to the preparation of ground-truth specifications, we inherit existing test cases in the original programs if there are or employ experts to manually craft them otherwise. The experts are required to cover as many branches as possible with the written test cases. False specifications (those in Section~\ref{sec:task_judge} and Section~\ref{sec:task_select}) are generated to test whether the test cases can disprove them, and more test cases should be crafted if they failed. Each program receives, on average, 15.10 test cases, with an average branch coverage of 95.54\% and line coverage of 93.34\%. 




\vspace{-2mm}
\subsection{Studied LLMs} \label{sec:subject_llms}

In this work, we selected six state-of-the-art LLMs for evaluation.
\vspace{-1mm}
\begin{itemize} [leftmargin=*]
    \item \textbf{GPT-3.5-Turbo}~\cite{openai2023gpt}, a general LLM developed by OpenAI~\cite{OpenAI}. We adopt \texttt{gpt-3.5-turbo-0125}, the latest model in the GPT-3.5 family during the experiment, for evaluation.
    \item \textbf{GPT-4-Turbo}~\cite{achiam2023gpt}, an improved general LLM developed by OpenAI, with leading performance among all general LLMs currently.
    \item \textbf{Llama 2}~\cite{touvron2023llama}, an open-source general LLM developed by Meta AI~\cite{MetaAI}. We adopt \texttt{Llama-2-7b-chat} for evaluation.
    \item \textbf{CodeLlama}~\cite{roziere2023code}, an open-source code LLM based on the architecture of Llama 2. \texttt{CodeLlama-7b-Instruct} is adopted.
    \item \textbf{DeepSeek-Coder}~\cite{guo2024deepseek}, an open-source code LLM employing the same architecture as DeepSeek LLM~\cite{bi2024deepseek}. We adopt the version \texttt{deepseek-coder-6.7b-instruct} for evaluation.
    \item \textbf{Magicoder}~\cite{wei2023magicoder}, an open-source code LLM fine-tuned with OSS-Instruct, a novel high-quality instruction tuning technique. The version employed in the evaluation is \texttt{Magicoder-S-DS-6.7B}.
\end{itemize}
The selected models have been applied to a wide variety of code-related tasks in existing works~\cite{chen2021evaluating,liu2023your,deng2023large}. They also exhibit diversities over multiple dimensions, ranging from general LLMs to code LLMs, open-source models to closed-source models, and foundation models to fine-tuned models. Due to the computing resource limit, only 7B versions of open-source LLMs are selected.

\vspace{-2mm}
\subsection{Prompt Format and Model Configurations} \label{sec:prompt}

We generally adopt a unified prompt structure for each task and all the subject LLMs in the experiments. The structure mainly consists of four components. Initially, a system message is appended to the context to inform the model of its basic role. The few-shot prompting technique is then adopted, where several pairs of example requests and replies are provided to the LLM, assisting the LLM in learning the task requirements and desired output format. \lz{In our experiments, a 2-shot setting is adopted. The few-shot examples are collected from the official tutorial of JML~\cite{opemjml2025tutorial}, including example specifications that cover the fundamental syntax and semantics of JML, enough for LLMs to understand and accomplish the tasks.} Afterward, a description of the target task (i.e., the current task that the LLM is expected to answer) is appended to the context. Up to this point, the context constitutes a complete query and is sent to the LLM, which learns necessary information from the context and completes the task within its reply. 

For the open-source models mentioned in Section~\ref{sec:subject_llms}, we deploy a local server to execute the models and process relevant queries, running on a Linux Server equipped with one H800 GPU with 80GB memory. The source code and model weights of all open-source models can be accessed at corresponding HuggingFace~\cite{huggingface2024huggingface} repositories. For GPT-3.5 and GPT-4, we access them through the API provided by OpenAI~\cite{openai2024api}. Following the settings of previous works~\cite{du2024evaluating,liu2024codemind}, we adopt greedy decoding~\cite{chen2015discrete} for all subject LLMs to make the results reproducible. Generally, the models are configured as \texttt{top\_k=1} and \texttt{temperature=0} under greedy decoding. Other parameters are kept consistent with default settings.


%% file: sections/evaluation.tex
\vspace{-1mm}

\begin{table*}[ht]
\caption{Performance of all models on each task and perturbation category. \textbf{Acc.}: Average Accuracy, \textbf{AvgPrec.}: Average Precision, \textbf{AvgRec}: Average Recall in percentages. \textbf{\#Pass}: the number of programs handled with all test cases passed. } \label{tab:overall}
\vspace{-3mm}
\centering
\resizebox{1\textwidth}{!}{
\begin{tabular}{@{}c|c|ccccccc|c|c|ccccccc@{}}
\toprule
\multirow{2}{*}{Category} & \multirow{2}{*}{LLM} & Judgement & Selection & \multicolumn{2}{c}{Infilling} & \multicolumn{3}{c|}{Generation} & \multirow{2}{*}{Category} & \multirow{2}{*}{LLM} & Judgement & Selection & \multicolumn{2}{c}{Infilling} & \multicolumn{3}{c}{Generation} \\ \cmidrule(lr){3-4} \cmidrule(lr){5-6} \cmidrule(lr){7-9} \cmidrule(l){12-13} \cmidrule(l){14-15} \cmidrule(l){16-18} 
 &  & Acc. & Acc. & Acc. & \#Pass & Prec. & Rec. & \#Pass &  &  & Acc. & Acc. & Acc. & \#Pass & Prec. & Rec. & \#Pass \\ \midrule
\multirow{6}{*}{Original} & GPT-3.5 & 64.71 & 67.65 & 34.31 & 70 & 73.22 & 25.35 & 106 & \multirow{6}{*}{\begin{tabular}[c]{@{}c@{}}Def-use\\ Break\end{tabular}} & GPT-3.5 & 67.65 & 61.27 & 36.76 & 75 & 70.51 & 29.82 & 98 \\
 & GPT-4 & \textbf{81.37} & \textbf{89.71} & \textbf{41.18} & \textbf{84} & \textbf{83.37} & \textbf{34.17} & \textbf{135} &  & GPT-4 & \textbf{78.92} & \textbf{86.27} & \textbf{40.20} & \textbf{82} & \textbf{77.81} & \textbf{46.03} & \textbf{112} \\
 & Llama & 50.00 & 30.88 & 10.78 & 22 & 29.65 & 4.07 & 24 &  & Llama & 54.41 & 24.02 & 6.86 & 14 & 24.68 & 3.26 & 20 \\
 & CodeLlama & 57.84 & 31.37 & 21.08 & 43 & 55.22 & 10.77 & 74 &  & CodeLlama & 62.25 & 28.43 & 19.61 & 40 & 46.01 & 10.77 & 55 \\
 & Deepseek-coder & 57.35 & 61.27 & 29.90 & 61 & 61.56 & 17.39 & 75 &  & Deepseek-coder & 60.29 & 59.80 & 24.51 & 50 & 61.80 & 23.38 & 72 \\
 & Magicoder & 44.12 & 39.71 & 29.41 & 60 & 64.16 & 13.72 & 81 &  & Magicoder & 53.43 & 45.59 & 29.41 & 60 & 55.53 & 19.76 & 60 \\ \midrule
\multirow{6}{*}{\begin{tabular}[c]{@{}c@{}}If-else\\ Flip\end{tabular}} & GPT-3.5 & 68.14 & 65.20 & 36.27 & 74 & 66.99 & 34.61 & 97 & \multirow{6}{*}{\begin{tabular}[c]{@{}c@{}}Independent\\ Swap\end{tabular}} & GPT-3.5 & 66.18 & 66.18 & 37.75 & 77 & 75.32 & 38.01 & 112 \\
 & GPT-4 & \textbf{77.94} & \textbf{83.82} & \textbf{40.69} & \textbf{83} & \textbf{79.16} & \textbf{48.26} & \textbf{115} &  & GPT-4 & \textbf{79.90} & \textbf{86.27} & \textbf{41.67} & \textbf{85} & \textbf{77.69} & \textbf{47.55} & \textbf{115} \\
 & Llama & 49.51 & 24.51 & 12.25 & 25 & 33.22 & 7.31 & 25 &  & Llama & 48.53 & 22.55 & 11.27 & 23 & 31.63 & 5.74 & 26 \\
 & CodeLlama & 62.25 & 31.37 & 17.65 & 36 & 49.00 & 16.36 & 56 &  & CodeLlama & 64.22 & 31.86 & 20.59 & 42 & 48.39 & 14.84 & 54 \\
 & Deepseek-coder & 61.27 & 57.35 & 27.45 & 56 & 63.26 & 27.27 & 79 &  & Deepseek-coder & 55.88 & 59.80 & 30.88 & 63 & 63.43 & 25.62 & 78 \\
 & Magicoder & 55.88 & 50.00 & 30.39 & 62 & 55.88 & 18.61 & 62 &  & Magicoder & 58.82 & 43.14 & 34.80 & 71 & 57.50 & 23.18 & 68 \\ \midrule
\multirow{6}{*}{\begin{tabular}[c]{@{}c@{}}Name\\ Random\end{tabular}} & GPT-3.5 & 63.24 & 54.90 & 35.78 & 73 & 71.65 & 38.14 & 103 & \multirow{6}{*}{\begin{tabular}[c]{@{}c@{}}Name\\ Shuffle\end{tabular}} & GPT-3.5 & 68.14 & 66.67 & 33.82 & 69 & 68.27 & 29.91 & 97 \\
 & GPT-4 & \textbf{76.96} & \textbf{81.37} & 35.29 & 72 & \textbf{81.18} & \textbf{50.65} & \textbf{124} &  & GPT-4 & \textbf{76.96} & \textbf{84.31} & \textbf{35.78} & \textbf{73} & \textbf{77.39} & \textbf{45.69} & \textbf{111} \\
 & Llama & 56.37 & 25.49 & 10.29 & 21 & 28.73 & 5.70 & 25 &  & Llama & 49.02 & 26.96 & 8.82 & 18 & 32.64 & 5.09 & 31 \\
 & CodeLlama & 61.27 & 24.51 & 19.61 & 40 & 49.10 & 12.75 & 65 &  & CodeLlama & 61.27 & 30.88 & 22.06 & 45 & 43.70 & 13.06 & 48 \\
 & Deepseek-coder & 55.88 & 61.27 & 27.45 & 56 & 61.71 & 26.86 & 69 &  & Deepseek-coder & 59.80 & 59.31 & 25.49 & 52 & 61.40 & 27.58 & 70 \\
 & Magicoder & 60.29 & 48.04 & \textbf{37.25} & \textbf{76} & 52.04 & 22.24 & 63 &  & Magicoder & 63.73 & 46.08 & 28.92 & 59 & 54.17 & 21.37 & 63 \\ \bottomrule
\end{tabular}
}
\vspace{-4mm}
\end{table*}

\section{Experimental Results} \label{sec:evaluation}

Our experiments aim to answer the following research questions:
\begin{itemize} [leftmargin=*]
    \item \textbf{RQ1:} How is the general performance of each LLM evaluated on the tasks related to formal specifications? 
    \item \textbf{RQ2:} How does few-shot prompting strategy affect the performance of LLMs in the tasks involved?
    \item \textbf{RQ3:} How is the performance variance of LLMs when confronted with semantic-preserving perturbations? 
\end{itemize}

\vspace{-2mm}
\subsection{RQ1: Overall Performance} \label{sec:RQ1}

Table~\ref{tab:overall} shows the overall performance of each model on four tasks for all versions of programs (including original and five perturbed variants). 
\vspace{-1mm}
\subsubsection{Comparison Between Tasks} \label{sec:RQ1-1}
For different tasks involved in the experiments, the models showcase different levels of fulfillment. We take the \textit{original} category as the example for illustration. For the \textit{Judgement} task, most models can achieve an accuracy below 65\% except the 81.37\% of GPT-4. Given the high similarity between this task and the model checking technique, the models may lack the symbolic reasoning capabilities required for model checking. For the \textit{Selection} task, the accuracy of most models either shows some decline compared to that of \textit{Judgement} or has no significant variance. When it comes to \textit{Infilling}, we can witness a significant drop in the performance of all models. As for \textit{Generation}, the average precision ranges from 29.65\% of Llama to 83.37\% of GPT-4, whereas the average recall is much lower, varying between 4.07\% of Llama and 34.17\% of GPT-4. The low recall of all LLMs suggests that they can only generate limited specifications with strength comparable to the ground truth, indicating a relatively weak ability to summarize full program semantics into formal specifications. The other five categories also display similar patterns.

According to their performance, the most straightforward task involved in the experiment is \textit{Judgement}, followed by \textit{Selection}, then \textit{Generation}, whereas the most challenging task is \textit{Infilling}. It is an interesting finding, and to our intuition, the root cause of this phenomenon lies in the different aspects of ability that are examined by different tasks. \textit{Judgement} and \textit{Selection} are mainly about \textit{reading and understanding} (i.e., "input") specifications, where LLMs try to comprehend given specifications and give relevant conclusions. On the contrary, \textit{Generation} is about \textit{writing and expressing} (i.e., "output") specifications, where LLMs can \textit{freely} articulate code semantics they understand in the form of specifications, and the specifications written may not be strong. Interestingly, \textit{Infilling} lies somewhere in between, where LLMs \textit{write} partial specifications but cannot do it at any will. They have to generate results within the structure defined by the target specification to fill in, which is usually strong with sufficient semantic information. To this end, they have to \textit{read and understand} the given specification first, which may be difficult for them, and then, based on their understanding, figure out the \textit{desired} answers. Consequently, the \textit{Infilling} task examines \textit{both} of the two aspects mentioned above (i.e., "input" \textit{and} "output") simultaneously, making it the most difficult task of all.

\begin{table*}[!t]
\caption{Performance of all models under different few-shot settings on the \textit{Original} category. Config: The adopted few-shot configuration. Acc.: Average Accuracy, Prec.: Average Precision, Rec.: Average Recall (all in percentages).}
\label{tab:fewshot}
\vspace{-3mm}
\centering
\resizebox{1\linewidth}{!}{
\begin{tabular}{c|c|ccccc|c|c|ccccc|c|c|ccccc}
\toprule
\multirow{2}{*}{Model} & \multirow{2}{*}{Config} & Judgment & Selection & Infilling & \multicolumn{2}{c|}{Generation} & \multirow{2}{*}{Model} & \multirow{2}{*}{Config} & Judgment & Selection & Infilling & \multicolumn{2}{c|}{Generation} & \multirow{2}{*}{Model} & \multirow{2}{*}{Config} & Judgment & Selection & Infilling & \multicolumn{2}{c}{Generation} \\ \cmidrule(lr){3-4} \cmidrule(lr){5-5} \cmidrule(lr){6-7} \cmidrule(lr){10-11} \cmidrule(lr){12-12} \cmidrule(lr){13-14} \cmidrule(lr){17-18} \cmidrule(lr){19-19} \cmidrule(lr){20-21} 
 &  & Acc. & Acc. & Acc. & Prec. & Rec. &  &  & Acc. & Acc. & Acc. & Prec. & Rec. &  &  & Acc. & Acc. & Acc. & Prec. & Rec. \\ \midrule
\multirow{3}{*}{GPT-3.5} & 0-shot & 62.25 & 25.49 & 29.90 & 34.83 & 7.30 & \multirow{3}{*}{GPT-4} & 0-shot & 77.94 & 25.00 & 36.76 & \textbf{83.64} & 32.73 & \multirow{3}{*}{Llama} & 0-shot & 42.65 & 25.00 & \textbf{20.10} & 0.98 & 0.00 \\
 & 1-shot & 64.71 & 62.75 & 26.96 & 59.66 & 15.87 &  & 1-shot & 81.37 & 85.29 & 37.75 & 79.56 & 31.23 &  & 1-shot & \textbf{56.86} & 26.96 & 14.22 & \textbf{25.63} & 2.00 \\
 & 2-shot & \textbf{64.71} & \textbf{67.65} & \textbf{34.31} & \textbf{73.22} & \textbf{25.35} &  & 2-shot & \textbf{81.37} & \textbf{89.71} & \textbf{41.18} & 83.37 & \textbf{34.17} &  & 2-shot & 50.00 & \textbf{30.88} & 14.22 & 24.01 & \textbf{15.88} \\ \midrule
\multirow{3}{*}{CodeLlama} & 0-shot & 52.45 & 23.53 & 3.43 & 6.21 & 1.82 & \multirow{3}{*}{\begin{tabular}[c]{@{}c@{}}Deepseek\\ Coder\end{tabular}} & 0-shot & 53.43 & 25.00 & 14.22 & 63.58 & 6.76 & \multirow{3}{*}{Magicoder} & 0-shot & 55.39 & 23.04 & 14.71 & \textbf{69.90} & 5.71 \\
 & 1-shot & 54.22 & 27.94 & 14.12 & 38.61 & 10.37 &  & 1-shot & 55.88 & 51.96 & 28.92 & \textbf{61.91} & 14.40 &  & 1-shot & \textbf{62.75} & \textbf{42.16} & 18.14 & 48.40 & 9.48 \\
 & 2-shot & \textbf{57.84} & \textbf{31.37} & \textbf{21.08} & \textbf{55.22} & \textbf{10.77} &  & 2-shot & \textbf{57.35} & \textbf{61.27} & \textbf{29.90} & 61.56 & \textbf{17.39} &  & 2-shot & 44.12 & 39.71 & \textbf{29.41} & 64.16 & \textbf{13.72} \\ \bottomrule
\end{tabular}
}
\vspace{-3mm}
\end{table*}

\vspace{-2mm}
\subsubsection{Comparison Between Models} \label{sec:RQ1-2}
Different models are specialized in different tasks. Regarding \textit{Judgement} and \textit{Selection}, GPT-4 outperforms all other models significantly with leading accuracy across all categories, followed by GPT-3.5. The three code LLMs involved in the experiments yield a relatively low accuracy. As described in Section~\ref{sec:RQ1-1}, \textit{Judgement} and \textit{Selection} are mainly about \textit{reading and understanding} specifications. This indicates that GPT-4 and GPT-3.5 showcase impressive ability regarding the comprehension of specifications. In terms of \textit{Infilling}, the gap between GPT-4 and GPT-3.5 is significantly narrower as task difficulty rises. Notably, Magicoder demonstrates the most outstanding performance among the open-source models, which even surpasses the GPTs in the \textit{Name Random} category. This indicates the possibility for open-source models to achieve comparable or even better performance compared to the closed-source GPT models with a much smaller model scale, under the assistance of fine-tuning with high-quality code data corpora. In the \textit{Generation} task, the models' ranking remains largely consistent, except for DeepSeek-Coder emerging as the top-performing model among the open-source ones.

\begin{figure}[!t]
    \centering
    \includegraphics[width=0.41\textwidth]{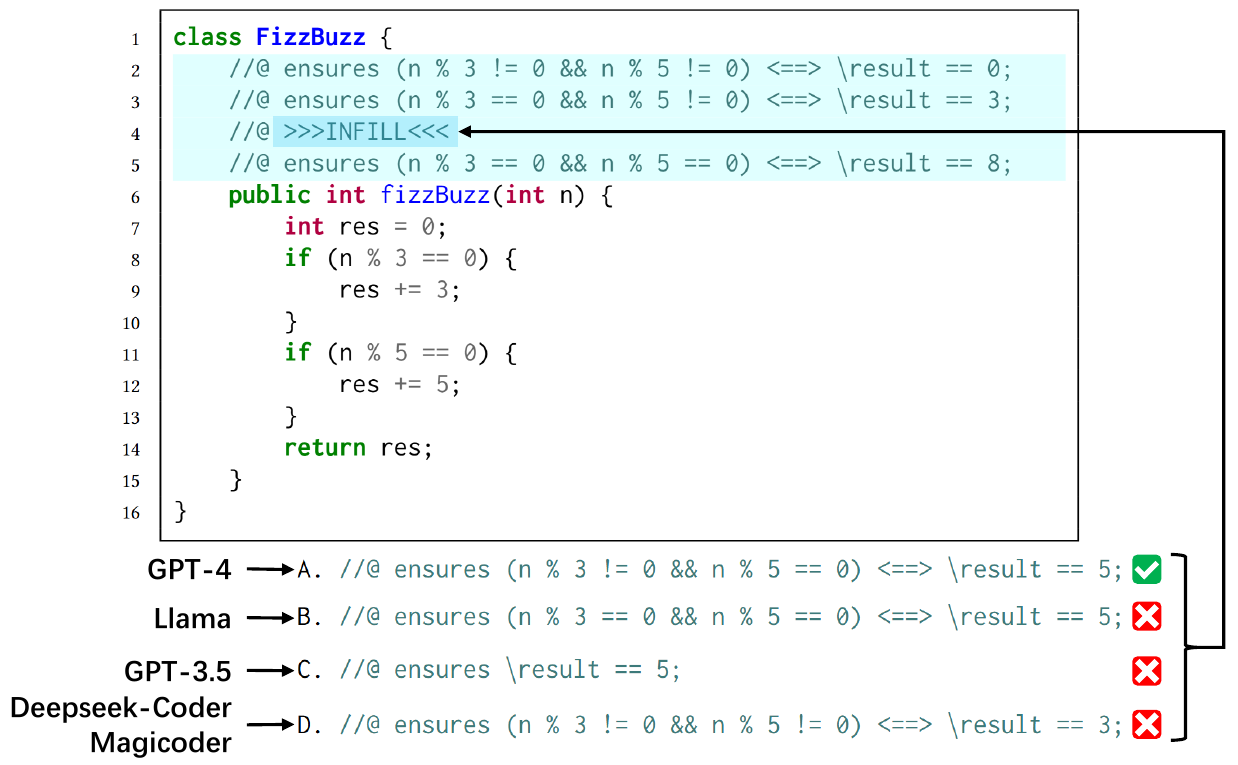}
    \vspace{-3mm}
    \caption{The selection task for program \texttt{FizzBuzz}, the correct answer, and the options chosen by each model.}
    \label{fig:case_study}
    \vspace{-4mm}
\end{figure}

\noindent\textbf{New Findings:} \lz{As is described in Section~\ref{sec:RQ1-1}, the performance of models such as CodeLlama on \textit{Selection} task degrades notably compared to \textit{Judgement}. Since selection tasks require models to distinguish correct specifications from incorrect specifications that are highly similar, this phenomenon is possibly because the incorrect options can mislead the models, making them fail to distinguish the difference.} We further illustrate this finding with the selection task for program \texttt{FizzBuzz} shown in Fig.~\ref{fig:case_study}. The program determines its return value (denoted as \texttt{\textbackslash result} in JML) according to whether the input \texttt{n} is divisible by 3 and 5. Specifically, it returns 5 if and only if \texttt{n} is divisible by 5 but not 3, so the correct option is A. Option B and D are highly similar to A with only a few operators mutated. As marked in Fig.~\ref{fig:case_study}, all models provided incorrect answers except GPT-4. CodeLlama made up its own answer rather than choosing among the options given, which is also considered to be failed. There are two possible explanations for this phenomenon. First, for programs involving multiple (sometimes nested) control flow structures such as branches and loops, even though models can clarify the conditions and behaviors of individual structures separately, they may still struggle to synthesize their overall semantics, leading to incorrect comprehension or generation of formal specifications. \lz{59.8\% programs in our dataset involve multiple control flow structures, posing notable challenges to the models' reasoning and synthesis capabilities.} Second, models may be misled by information of uncertain correctness (e.g., the wrong options in the \textit{Selection} task) within the prompt, especially when correct and incorrect information is blended with highly similar syntactical structures and semantic contents. \lz{This is a novel intuition compared to those of prior works including CRUXEval~\cite{gu2024cruxeval} and REval~\cite{chen2024evaluating} as their tasks do not involve discrimination between correct and incorrect information.}

\vspace{-1mm}
\begin{tcolorbox}
\vspace{-2mm}
\noindent \textbf{Answer to RQ1: } LLMs showcase a limited ability to specify program semantics with formal specifications. Furthermore, \lz{LLMs are relatively weak at synthesizing together the semantics of multiple control-flow structures. Extraneous information with unknown correctness in prompts also poses a notable negative impact on the models' reasoning abilities, especially when correct and incorrect information are syntactically and semantically similar.}
\vspace{-2mm}
\end{tcolorbox}
\vspace{-2mm}

\vspace{-1mm}
\subsection{RQ2: Effectiveness of the Few-shot Setting} \label{sec:RQ2}

As is described in Section~\ref{sec:prompt}, we adopted few-shot prompts for all models. To investigate the effect of few-shot prompts on the models' comprehension abilities, and validate the effectiveness of our experimental design, we conduct further experiments by setting up 0-shot and 1-shot configurations in comparison with the 2-shot setting adopted in Section~\ref{sec:RQ1}.

Table~\ref{tab:fewshot} shows the performance of all models under different few-shot settings. Generally, the effect of few-shot learning is obvious, with different degrees of improvement on most tasks across all models. For \textit{Judgement}, the improvement of few-shot prompting is relatively modest, bringing less than 10\% improvement for all models. For \textit{Selection}, dramatic improvement can be observed on GPT-3.5, GPT-4, and Deepseek-Coder. As for \textit{Infilling}, the 2-shot setting can increase the accuracy of open-source models from around 10\% to around 30\% compared to 0-shot. For \textit{Generation}, the 2-shot configuration significantly improves the precision of GPT-3.5, Llama, and CodeLlama, while also enhancing the recall of all models to some extent. \lz{We did not conduct further experiments on configurations providing more few-shot examples due to frequent context length overflows and GPU memory shortages (with only one H800 GPU for experiments) caused by such configurations. Since we aim to evaluate model performance under a consistent setting to avoid bias rather than study the best setting for each model, the 2-shot setting is effective and fair.} 
Some models exhibit slight performance regression, such as the precision of GPT-4 in \textit{Generation}. This is due to the model's tendency to generate additional content to facilitate reasoning when no examples are provided to rectify its output.
Since some models may not have been previously trained on datasets containing extensive specifications, few-shot learning can provide them with the basic syntax and semantics of specification languages (i.e., JML in our experiments) within the context, equipping them with the fundamental knowledge needed to accomplish the experimental tasks. For these models, few-shot learning can significantly enhance their performance. For models already equipped with relevant knowledge, the assistance from few-shot learning may be relatively limited, but the provided few-shot examples can still achieve some improvements by familiarizing the models with the task content and rectifying their output format.

\vspace{-1mm}
\begin{tcolorbox}
\vspace{-2mm}
\noindent \textbf{Answer to RQ2: } Few-shot prompting can effectively improve the performance of all models across all tasks, especially when the models are not equipped with the basic knowledge of the specification language involved, where few-shot examples effectively help models learn specification syntax and semantics, validating the effectiveness of the 2-shot configurations in our experimental settings.
\vspace{-2mm}
\end{tcolorbox}
\vspace{-2mm}

\begin{table*}[!t]
\caption{Impact of all types of perturbations concerning different models and tasks. \textbf{DUB}: Def-use Break. \textbf{IEF}: If-else Flip. \textbf{IDS}: Independent Swap. \textbf{NMR}: Name Random. \textbf{NMS}: Name Shuffle. $J_{jud}$: Jaccard Distance for Judgement. $J_{sel}$: Jaccard Distance for Selection. $J_{inf}$: Jaccard Distance for Infilling. $v_{prec}$: Average Variance of Precision. $v_{rec}$: Average Variance of Recall.}
\label{tab:counterfactual}
\vspace{-3mm}
\centering
\resizebox{1\linewidth}{!}{
\begin{tabular}{c|c|ccccc|c|c|c|ccccc|c|c|cccccc|c}
\toprule
Model & Metric & DUB & IEF & IDS & NMR & NMS & Avg. & Model & Metric & DUB & IEF & IDS & NMR & NMS & Avg. & Model & Metric & DUB & IEF & IDS & NMR & NMS & Avg. \\ \midrule
\multirow{6}{*}{GPT-4} & $J_{jud}$ & 0.153 & 0.110 & 0.065 & 0.144 & 0.165 & \multirow{6}{*}{0.183} & \multirow{6}{*}{GPT-3.5} & $J_{jud}$ & 0.323 & 0.337 & 0.362 & 0.358 & 0.348 & \multirow{6}{*}{0.307} & \multirow{6}{*}{Llama} & \multicolumn{1}{c|}{$J_{jud}$} & 0.457 & 0.496 & 0.511 &  0.534 & 0.493 & \multirow{6}{*}{0.573} \\
 & $J_{sel}$ & 0.120 & 0.107 & 0.101 & 0.163 & 0.141 &  &  & $J_{sel}$ & 0.303 & 0.263 & 0.239 & 0.366 & 0.277 &  &  & \multicolumn{1}{c|}{$J_{sel}$} & 0.845 & 0.798 & 0.840 & 0.827 & 0.854 &  \\
 & $J_{inf}$ & 0.114 & 0.058 & 0.057 & 0.165 & 0.153 &  &  & $J_{inf}$ & 0.188 & 0.130 & 0.163 & 0.298 & 0.325 &  &  & \multicolumn{1}{c|}{$J_{inf}$} & 0.875 & 0.694 & 0.548 & 0.806 & 0.857 &  \\
 & $v_{prec}$ & 0.186 & 0.126 & 0.179 & 0.152 & 0.174 &  &  & $v_{prec}$ & 0.240 & 0.238 & 0.205 & 0.227 & 0.226 &  &  & \multicolumn{1}{c|}{$v_{prec}$} & 0.318 & 0.289 & 0.300 & 0.288 & 0.300 &  \\
 & $v_{rec}$ & 0.362 & 0.348 & 0.399 & 0.421 & 0.404 &  &  & $v_{rec}$ & 0.463 & 0.462 & 0.407 & 0.483 & 0.443 &  &  & \multicolumn{1}{c|}{$v_{rec}$} & 0.513 & 0.431 & 0.461 & 0.482 & 0.510 &  \\ \cmidrule{2-7} \cmidrule{10-15} \cmidrule{18-23}
 & Avg. & 0.187 & 0.150 & 0.160 & \textbf{0.209} & 0.207 &  &  & Avg. & 0.303 & 0.286 & 0.275 & \textbf{0.346} & 0.324 &  &  & \multicolumn{1}{c|}{Avg.} & 0.602 & 0.542 & 0.532 & 0.587 & \textbf{0.603} &  \\ \midrule
\multirow{6}{*}{CodeLlama} & $J_{jud}$ & 0.275 & 0.262 & 0.271 & \textbf{0.324} & 0.213 & \multirow{6}{*}{0.391} & \multirow{6}{*}{\begin{tabular}[c]{@{}c@{}}Deepseek\\ Coder\end{tabular}} & $J_{jud}$ & 0.389 & 0.376 & 0.362 & 0.396 & 0.374 & \multirow{6}{*}{0.340} & \multirow{6}{*}{Magicoder} & \multicolumn{1}{c|}{$J_{jud}$} & 0.526 & 0.563 & 0.521 & 0.521 & 0.523 & \multirow{6}{*}{0.496} \\
 & $J_{sel}$ & 0.373 & 0.222 & 0.303 & 0.459 & 0.329 &  &  & $J_{sel}$ & 0.236 & 0.284 & 0.210 & 0.252 & 0.230 &  &  & \multicolumn{1}{c|}{$J_{sel}$} & 0.597 & 0.524 & 0.580 & 0.579 & 0.529 &  \\
 & $J_{inf}$ & 0.544 & 0.451 & 0.455 & 0.617 & 0.533 &  &  & $J_{inf}$ & 0.318 & 0.254 & 0.203 & 0.375 & 0.338 &  &  & \multicolumn{1}{c|}{$J_{inf}$} & 0.621 & 0.494 & 0.422 & 0.583 & 0.616 &  \\
 & $v_{prec}$ & 0.408 & 0.349 & 0.364 & 0.357 & 0.379 &  &  & $v_{prec}$ & 0.318 & 0.289 & 0.300 & 0.288 & 0.300 &  &  & \multicolumn{1}{c|}{$v_{prec}$} & 0.376 & 0.325 & 0.337 & 0.335 & 0.323 &  \\
 & $v_{rec}$ & 0.416 & 0.470 & 0.407 & 0.465 & 0.521 &  &  & $v_{rec}$ & 0.513 & 0.431 & 0.461 & 0.482 & 0.510 &  &  & \multicolumn{1}{c|}{$v_{rec}$} & 0.477 & 0.491 & 0.482 & 0.513 & 0.548 &  \\ \cmidrule{2-7} \cmidrule{10-15} \cmidrule{18-23}
 & Avg. & 0.403 & 0.351 & 0.360 & \textbf{0.444} & 0.395 &  &  & Avg. & 0.355 & 0.327 & 0.307 & \textbf{0.358} & 0.351 &  &  & \multicolumn{1}{c|}{Avg.} & \textbf{0.519} & 0.480 & 0.468 & 0.506 & 0.508 &  \\ \bottomrule
\end{tabular}
}
\vspace{-2mm}
\end{table*}

\vspace{-1mm}
\subsection{RQ3: Counterfactual Analysis} \label{sec:RQ3}

The Jaccard Distances and Average Variances of different models on different tasks are listed in Table~\ref{tab:counterfactual}, both reflecting the perturbations' impact as described in Section~\ref{sec:counterfactual}. A higher value of these metrics indicates a higher impact imposed by the perturbation. Generally, all models exhibit performance variance to different degrees, indicating that the semantics LLMs learn are interfered with by the perturbations. Across all models, GPT-4 showcases minimal performance variance against all types of perturbations, with an average impact of 0.183, followed by GPT-3.5, with a slightly higher performance variance of 0.307. Among the open-source models, Deepseek-Coder exhibits the strongest consistency, with the average variance being 0.340. Other models demonstrate a higher performance variance and more sensitivity to perturbations. Also, the impact of perturbations varies among different tasks, tending to be stronger in the tasks where the models underperform. For instance, as described in Section~\ref{sec:RQ1}, GPT-4 significantly outperforms all the other models in most circumstances, where it also possesses the lowest performance variance, showing the best consistency over all types of perturbations. Also, all models yield relatively lower \textit{Recall} in the \textit{Generation} task, and correspondingly, the $v_{rec}$ metrics of most models are also significantly higher than others. 

\begin{figure}[!t]
    \centering
    \includegraphics[width=0.47\textwidth]{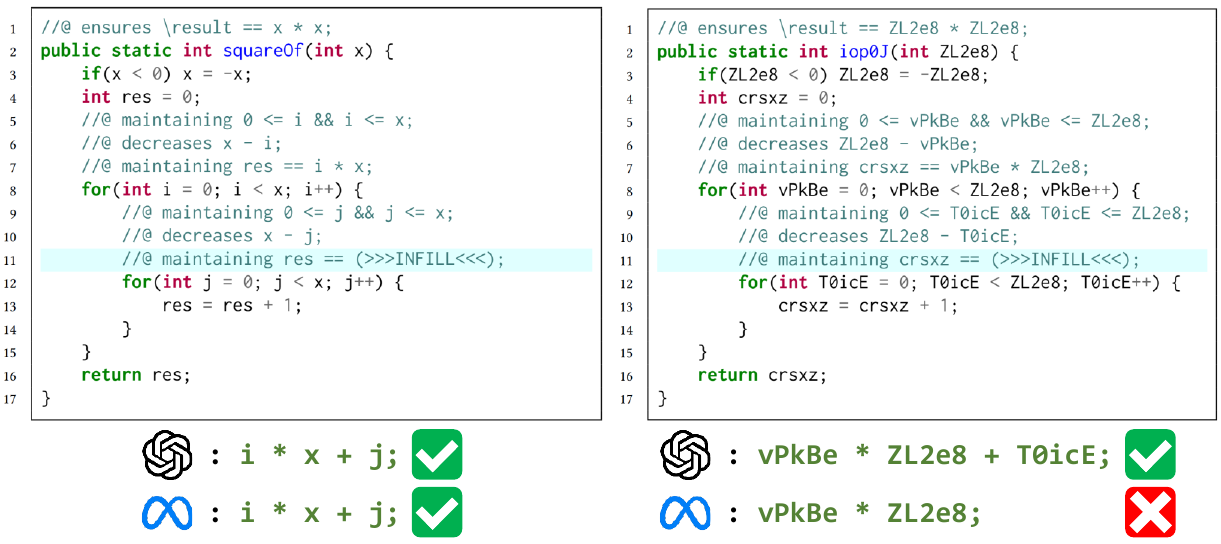}
    \vspace{-3mm}
    \caption{The infilling task for program \texttt{IntSquare} (both the \textit{Original} and the \textit{Name Random} category), along with the answers provided by GPT-3.5 and CodeLlama.}
    \label{fig:case_study2}
    \vspace{-4mm}
\end{figure}

\noindent\textbf{New Findings:} Concerning the average impact of \lz{the commonly used} perturbations, \textit{Name Random} demonstrated the most significant impact, with 4 out of the 6 models showing the most sensitivity to it, followed by \textit{Def-use Break} and \textit{Name shuffle}, which also pose a notable influence on the models' performance. The commonality among these three types of perturbations lies in their alteration of variable names, which disrupts the semantic information contained. To further illustrate this, two versions (\textit{Original} and \textit{Name Random} perturbed) of program \texttt{IntSquare} along with their \textit{Infilling} tasks are provided in Figure~\ref{fig:case_study2}. The models should reason about the constraints on a series of loop variables to complete the loop invariant on Line 11. For the \textit{Original} version, both GPT-3.5 and CodeLlama completed the task successfully. However, when it comes to the \textit{Name Random} perturbed version, CodeLlama failed to grasp the functionality of each variable effectively faced with the perturbed identifiers. It only replicated the loop invariant from Line 7, which coincidentally shares the same prefix with the target one, resulting in an incorrect answer. In contrast, GPT-3.5 was able to maintain the correct reasoning. \lz{As suggested by previous works~\cite{miceli2023larger,wang2023does,liu2023contrabert}, variable naming can affect model performance.} Intuitively, identifiers can carry semantic information such as variable functionalities, helping models in semantic comprehension. Irregular variable names reduce the semantic information that models can learn from them, causing fluctuations in model performance. Furthermore, this case reveals the tendency of LLMs to imitate the prompts given blindly when being unable to reason effectively, despite the potential inapplicability of these contents. Compared to these perturbations, \textit{If-else Flip} and \textit{Independent Swap} show a relatively lower impact since they have less interference with the approaches to acquiring program semantics. \lz{This is a novel finding compared to prior works~\cite{gu2024cruxeval,chen2024evaluating}, since they did not conduct fine-grained Counterfactual Analysis incorporating semantic-preserving perturbations for LLMs, through which we discover the implicit semantic information embedded in identifiers and the impact of perturbing them on model performance.}


\vspace{-1mm}
\begin{tcolorbox}
\vspace{-2mm}
\noindent \textbf{Answer to RQ3: } Perturbations can interfere with the semantics learned. Models tend to exhibit stronger sensitivity to perturbations in the tasks where they underperform. \lz{Among the semantic-preserving perturbations, the most influential ones are those that interfere with the identifiers. Since identifiers can carry semantic information, perturbations disrupting them can pose a greater impact on the models.}
\vspace{-2mm}
\end{tcolorbox}
\vspace{-2mm}

%% file: sections/discussion.tex
\vspace{-1mm}
\section{discussion} \label{sec:discussion}

\subsection{Implication}

\lz{The implications of the intuitions described in Section~\ref{sec:evaluation} cover multiple parties involved with LLMs. For \textit{LLM developers}, more refined datasets and fine-tuning techniques should be prepared to improve LLMs' resistance against perturbations and extraneous information. \textit{For LLM testers}, it is necessary to incorporate adversarial test cases with misleading information to thoroughly evaluate model robustness. \textit{For LLM-based tool developers}, applying validation and error-correction mechanisms on model output is crucial to improving tool reliability. For software developers utilizing LLMs, the model prompts should be cautiously constructed to avoid interference from extraneous information that may disrupt the models' reasoning.} Except for those directly associated areas, in terms of \textit{code completion and generation} with LLMs, specifications can assist the quality assurance of LLM-generated code. Concerning the \textit{software development} process, all types of specifications (including documents and comments) deserve attention since they improve code readability and maintainability. 

\vspace{-2mm}
\subsection{Greedy Decoding for Model Generation}

Following previous works~\cite{liu2023your,du2024evaluating,liu2024codemind}, we adopted greedy decoding (\texttt{top\_k=1} and \texttt{temperature=0}) for all LLMs in the experiment to guarantee the determinism in the sampling process of LLMs and the reproducibility of the final results. Nevertheless, some research within the same domain~\cite{gu2024cruxeval,chen2024evaluating} adopted random sampling for evaluation, which is also a choice. Currently, research has emerged to compare the performance of LLMs under greedy decoding and other decoding strategies. Cobbe et al.~\cite{cobbe2021training} and Hendrycks et al.~\cite{hendrycks2021measuring} show that greedy decoding is preferred when math reasoning problems are to solve. Song et al.~\cite{song2024good} claim that greedy decoding is generally more effective for most tasks, especially reasoning tasks and coding problems. Since the specification-related tasks involved in this work are generally about reasoning and formulating formal statements, greedy decoding tends to be the more suitable choice.

\vspace{-2mm}
\subsection{Threats to Validity and Limitations}

\noindent\textbf{Internal Threats.} First, experimental results may be affected by the prompts utilized to communicate with LLMs. We follow the prompt structure of prior work.~\cite{chen2024evaluating}. The effect of prompt techniques on the model performance will be studied in future works. Second, the base datasets, based on which we formulate the adapted benchmark, face potential data leakage when using the LLMs for evaluation. For programs in Frama-C-Problems~\cite{github2024framac} and SV-COMP~\cite{sosylab2024svcomp}, the ground-truth JML specifications involved are manually crafted by experts without risks of data leakage. As for SpecGenBench~\cite{ma2024specgen}, the dataset was publicly released in March 2024, whereas the versions of models we adopted were all released before this point in time. Thus, using our selected LLMs for evaluation is without data leakage issues. 
\lz{The last threat lies in the specification-related knowledge required by the experimental tasks, which some models may lack, potentially leading to under-performance in the tasks. To tackle this, we adopt the few-shot learning technique so that the models can learn the fundamental syntax and semantics of JML from the examples given, which is effective as is analyzed in Section~\ref{sec:RQ2}.}

\noindent\textbf{External Threats.} The external threat lies in the specification verifier. OpenJML~\cite{cok2011openjml} provides static verification and runtime checking for validation. Static verification requires sufficient supporting information (i.e., other strong specifications) and fails easily if not given enough, introducing serious bias. Therefore, runtime checking, with more stability and reliability, is adopted instead. 

\noindent\textbf{Limitations.} \lz{The main limitation lies in the limited size of our dataset due to the difficulties in manually crafting reliable specifications. To compensate for this, we design multiple tasks while performing various equivalent perturbations on the existing programs, thus increasing the volume of experiments, improving the diversity of experimental content, and enhancing the reliability of the experimental results. Our benchmark involves a total of 4896 tasks for each subject model evaluated, whereas the number is 3152 for REval. We plan to employ existing specification generation techniques~\cite{ma2024specgen,wen2024enchanting} to enlarge the benchmark in the future.}




%% file: sections/related.tex
\vspace{-1mm}
\section{Related work} \label{sec:related}

\noindent \textbf{LLM-based Specification Generation.} Recent works have been focusing on generating program specifications with LLMs. Janssen et al.~\cite{janssen2024can} investigated the ability of ChatGPT~\cite{ChatGPT} to infer loop invariants for C programs. Pei et al.~\cite{pei2023can} utilized fine-tuning to enhance the performance of LLMs on specification generation tasks. Ma et al.~\cite{ma2024specgen} combined LLM conversations with mutations to generate specifications for Java programs. Wei et al.~\cite{wen2024enchanting} adopted the program decomposition technique to assist the LLM-based specification generation process. For Rust code, Yang et al.~\cite{yang2024autoverus} utilized LLM agent networks to generate proofs automatically, whereas Chen et al.~\cite{chen2024automated} established a self-evolving cycle incorporating data synthesis and fine-tuning to enhance LLMs' generation ability. Compared to these works, we aim to evaluate LLMs' code comprehension abilities through specification-related tasks.

\noindent \textbf{LLM Benchmarks.} Some popular code generation benchmarks such as HumanEval~\cite{chen2021evaluating}, MBPP~\cite{austin2021program} and SWE-bench~\cite{jimenez2023swe} have been proposed to evaluate LLMs' coding capabilities. Other benchmarks targeting code comprehension with different downstream tasks have also been established. CRUXEval~\cite{gu2024cruxeval} is a dataset focusing on input/output prediction tasks for simple Python functions to evaluate the code reasoning and execution abilities of LLMs. Inspired by CRUXEval, REval~\cite{chen2024evaluating} further assesses LLMS' code reasoning ability through four different tasks concerning dynamic execution prediction. Compared to CRUXEval and REval, our framework utilizes program specifications as a carrier of \textit{overall} program semantics to evaluate LLMs' general understanding on programs' essential behaviors and functionalities, rather than reasoning about program states within a single execution trace. This novel perspective leads us to new findings such as the deficiency of LLMs in synthesizing the semantics of multiple control flow structures as described in Section~\ref{sec:RQ1}. Additionally, some other works incorporating formal method tasks to evaluate LLM ability have been proposed. He et al.~\cite{he2024beyond} introduce an LLM maturity model by assessing LLMs' performance on post-condition generation tasks. Compared to this work, we focus on a wider variety of specification-related tasks and a broader spectrum of specification types (including pre-conditions, post-conditions and loop invariants) to conduct a more comprehensive evaluation. \lz{Cao et al.~\cite{cao2025informal} constructed an extensive dataset to investigate model performance improvements on formal methods tasks brought about by fine-tuning. Compared to this work, we focus on the program semantics learned by LLMs by assessing the formal specifications generated, while incorporating Counterfactual Analysis to study the performance variance of LLMs when faced with semantic-preserving perturbations, making further discoveries regarding the semantic information embedded within identifiers as described in Section~\ref{sec:RQ3}.}

%% file: sections/conclusion.tex
\vspace{-1mm}
\section{conclusion} \label{sec:conclusion}

We present \ourname, a novel black-box framework evaluating the code comprehension ability in Large Language Models via program specifications. Leveraging the feature that specifications cover abundant execution traces and rich semantic information, we adopt them as an effective representation of program semantics. Four specification-related tasks are designed to assess the performance of LLMs. Counterfactual Analysis is further designed to assess the sensitivity of LLMs to semantic-preserving perturbations. Experimental results show that LLMs showcase a limited ability to fully articulate program semantics with formal specifications, with performance inconsistency when confronted with perturbations.